\documentclass[conference]{IEEEtran}
\usepackage{cite}
\usepackage{amsmath,amssymb,amsfonts}
\usepackage{graphicx}
\usepackage{textcomp}
\usepackage{xcolor}
\usepackage{multirow}
\usepackage[normalem]{ulem}
\usepackage{epsfig, subfigure, amsmath, amssymb}
\usepackage{esvect}
\usepackage{optidef}
\usepackage[ruled,vlined]{algorithm2e}
\usepackage{balance}

\SetKwInput{KwInput}{Input}              
\SetKwInput{KwOutput}{Output}            

\usepackage{amsmath}
\usepackage{graphicx}
\usepackage[utf8]{inputenc}

\definecolor{aqua}{rgb}{0.7,0,0.7}

\makeatletter
\let\@authorsaddresses\@empty
\makeatother

\IEEEoverridecommandlockouts

\author{
Alvi Ataur Khalil\IEEEauthorrefmark{1}, Javier Franco\IEEEauthorrefmark{2}, Imtiaz Parvez\IEEEauthorrefmark{3}, Selcuk Uluagac\IEEEauthorrefmark{2}, Mohammad Ashiqur Rahman\IEEEauthorrefmark{1}\\
\IEEEauthorrefmark{1}Analytics for Cyber Defense (ACyD) Lab, Florida International University, USA\\
\IEEEauthorrefmark{2}Cyber-Physical Systems Security Lab (CSL), Florida International University, USA\\
\IEEEauthorrefmark{3}Department of Electrical and Computer Engineering, Florida International University, USA\\
\IEEEauthorrefmark{1}\{akhal042, marahman\}@fiu.edu, \IEEEauthorrefmark{2}\{jfran243, suluagac\}@fiu.edu, \IEEEauthorrefmark{3}iparv001@fiu.edu
}

\begin{document}

\title{A Literature Review on Blockchain-enabled Security and Operation of Cyber-Physical Systems}


\maketitle

\begin{abstract}


Blockchain has become a key technology in a plethora of application domains owing to its decentralized public nature. The cyber-physical systems (CPS) is one of the prominent application domains that leverage blockchain for myriad operations, where the Internet of Things (IoT) is utilized for data collection. Although some of the CPS problems can be solved by simply adopting blockchain for its secure and distributed nature, others require complex considerations for overcoming blockchain-imposed limitations while maintaining the core aspect of CPS. Even though a number of studies focus on either the utilization of blockchains for different CPS applications or the blockchain-enabled security of CPS, there is no comprehensive survey including both perspectives together. To fill this gap, we present a comprehensive overview of contemporary advancement in using blockchain for enhancing different CPS operations as well as improving CPS security. To the best of our knowledge, this is the first paper that presents an in-depth review of research on blockchain-enabled CPS operation and security.

\end{abstract}

\begin{IEEEkeywords}
Blockchain, Cyber-physical systems, Data security, Internet of Things
\end{IEEEkeywords}

\vspace{-5pt}
\section{Introduction}
\label{sec:introduction}\vspace{-5pt}

\begin{figure}[hbt!]
    \centering
    \includegraphics[width=0.8\columnwidth]{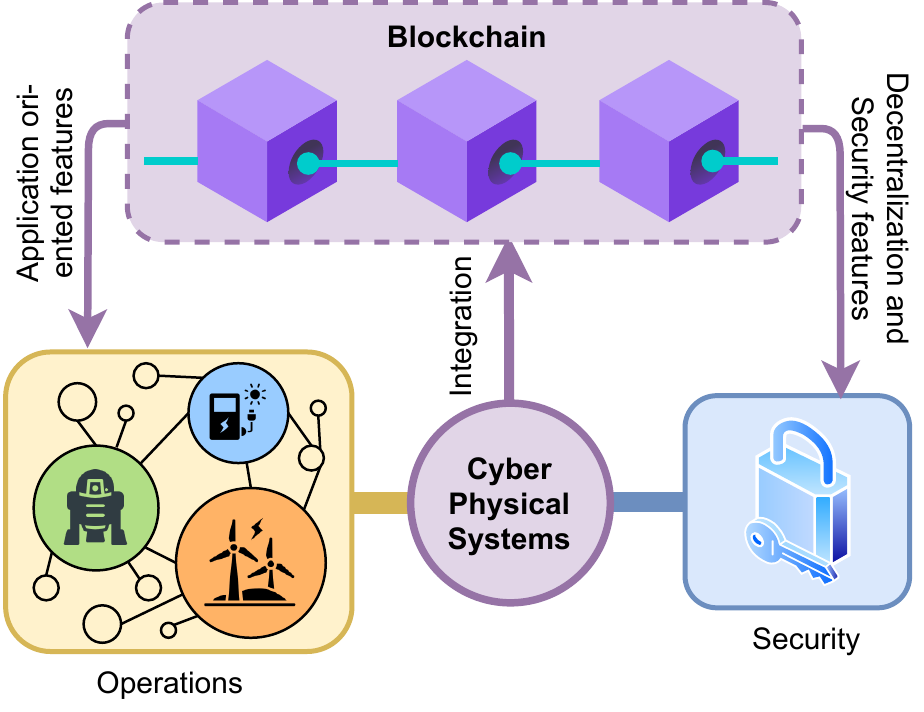}
    \vspace{-9pt}
    \caption{Blockchain enabling CPS operations and security.}
    \label{fig:bc_cps}
    \vspace{-5pt}
\end{figure}

Cyber-Physical Systems (CPS) have become essential for critical infrastructure worldwide, including water, energy, gas, healthcare, transportation, and smart grid systems. These systems include Internet of Things (IoT) devices that generate a massive volume of data, which they communicate to a centralized system. However, these devices have resource constraints for data storage, processing, and security measures, which pose significant challenges for the security and efficiency of CPS. As attackers are increasingly carrying out more directed attacks, CPS have become important targets to achieve maximum impact. A recent example of the far-reaching impacts of an attack on CPS is the recent Colonial Pipeline malware attack \cite{Padilla:2021}. This heightens the global importance of effective CPS security solutions.  As the number of these interconnected devices continues to grow, with an estimated 29.3 billion networked devices by 2023 \cite{Cisco:2020}, blockchain has emerged as a significant component in restructuring CPS systems for increased security and efficiency, as shown in Fig.~\ref{fig:bc_cps}.

Blockchain is regarded as one of the most important technologies that will bring about the next society transformation into the future~\cite{casino2019systematic}. Decentralization, immutability, distributed trust, increased security, smart contracts, digital currency, faster settlements, and minting are all properties of blockchain that can be utilized to address different challenges of CPS. To include in shared transactions with tamper-proof records, IoT devices/CPSs will be able to transfer the data to blockchain that is private in nature. Owing to the blockchain's distributed replication, diverse CPS data users can supply data from IoT sources without the requirement for core management and control systems. Each transaction may be verified by all the stakeholders belonging to the ecosystem of the CPS, avoiding disagreements and guaranteeing that each user is accountable for his particular parts in the entire transaction. 
Although solutions provided by blockchain are being adopted widely in the contemporary CPS domains, because of the different capabilities discussed, there are a lot of challenges in meeting the diverse requirements of different CPS applications. 

From the literature, it is observed that a lot of the blockchain-based security and operation surveys have been conducted in the CPS domain. However, the review focus was limited to specific considerations for either the operation or the security of CPS achieved through leveraging blockchain. In this work, we provide a detailed review of the research works conducted in the blockchain-enabled CPS domain from both the perspectives, and to the best of our knowledge, this is the first paper with this level of exhaustive overview of blockchain-enabled CPS.

The rest of the paper is organized as follows: We provide sufficient preliminary information in Section~\ref{sec:background}. The related works are discussed in Section~\ref{sec:related_works}. We discuss the literature related to Blockchain-enabled CPS in Section~\ref{sec:research}. We present a statistical analysis of the literature in Section~\ref{sec:discussion}. Lastly, we conclude the paper in Section~\ref{sec:conclusion}.


\vspace{-2pt}
\section{Background}\vspace{-3pt}
\label{sec:background}

In this section, we provide some introductory information regarding both CPSs and blockchain.

\vspace{-3pt}
\subsection{Cyber-Physical Systems}\vspace{-3pt}
The concept of CPS is based on systems that incorporate both cyber and physical systems to exchange data in real time. A CPS is a network of embedded systems consisting of sensors, aggregators, and actuators that are capable of monitoring and controlling real IoT-related processes and objects \cite{Yaacoub:2020}. CPS consists of the integration of sensing, networking, communication, control, and computation. 
\vspace{-3pt}
\subsection{Blockchain}\vspace{-3pt}
Blockchain is a decentralized and distributed method of recording and tracking digital interactions \cite{Shahid:2019}. Zhao \cite{Zhao:2021} describes how blockchain utilizes a chain-like data structure, which operates on a peer-to-peer network without a centralized trusted authority, and uses cryptography such as cryptographic hash and public-key cryptography. Each block contains various transactions, and blocks are chained together and have great redundancy. Therefore, if any blocks are altered or removed, this can easily be identified, and this also makes it very difficult to damage information on the blockchain. Furthermore, blockchain uses the Proof of Work (PoW) algorithm, which is used to validate transactions and create new blocks on the chain through solving a complex mathematical puzzle \cite{Zhao:2021}. 


\vspace{-2pt}
\section{Related Works}\vspace{-3pt}
\label{sec:related_works}

A vast number of review articles on blockchain-enabled CPS have been published, each covering a distinct component of this research methodology. Many of these surveys focus on CPS security, like Taylor et al. identified peer-reviewed literature regarding cyber security through blockchain by exploring various adopted blockchain security applications in~\cite{taylor2020systematic}. They highlighted the potentials of different research studies in the cybersecurity domain, even excluding the IoT, by blockchain applications. Gupta et al. offered a survey in~\cite{gupta2020smart} that is primarily concerned with the cybersecurity vulnerabilities of smart contracts in blockchain enabled CPS applications, where software code can be easily hacked by the adversarial users. They found that even complex designs of smart contracts fail to mitigate the security issues and accordingly they investigated Artificial Intelligence (AI) techniques for smart contract privacy protection. Keshk et al. \cite{Keshk:2021} provide a survey of current privacy-preserving techniques that are used to protect CPS systems and their data from cyber-attacks. They classify and explain privacy protection techniques, including blockchain. 

Others focus on control and operation of blockchain-based CPS. Zhao et al. dissected various blockchain-enabled CPS in terms of the operations and features utilized, and classified them according to the sensitivity and throughput in~\cite{Zhao:2021}. 
Kanhere addressed in~\cite{kanhere2020keynote} that, although a decentralized approach realizes the true potential of CPS taking the unique features into account, the application of blockchain for diverse CPS domains has its own complex challenges. Braeken et al. shed light into the technical and societal challenges, solutions and opportunities in various application domains combining the benefits of blockchain and cyberphysical system~\cite{braeken2020blockchain}. 
In~\cite{bodkhe2020survey}, Bodkhe et al. explored the state-of-the-art consensus mechanisms, highlighting their strengths as well as weaknesses in decentralized CPS applications, through a comprehensive analysis. They further present the gaps in existing surveys and propose a solution taxonomy of decentralized consensus mechanisms for various CPS applications. A holistic survey of different CPS application domains including smart grids, health-care systems, and industrial production processes leveraging blockchain for robustness and reliability, has been presented in~\cite{rathore2020survey}. They additionally provide a mathematical model for determining if a certain application may benefit from the blockchain. Finally, Dedeoglu et al. addressed in~\cite{dedeoglu2020journey} that high latency, low scalability and throughput, and computationally expensive consensus mechanisms greatly hinder the mass adoption of blockchain in the CPS application domain.  

Each of these studies sheds light on important considerations for the usage of blockchain in CPS. However, none of the existing studies provides a focus on the research trends in using blockchain for enhancing  CPS in  different  operations  as well as improving security of CPS.


\vspace{-2pt}
\section{Research Studies}\vspace{-3pt}
\label{sec:research}

In the following section, we classify recent studies by their focus. All the studies are certainly interrelated, and many could apply to several of the categories.  However, in classifying the studies, we highlight the key objectives identified by the authors in order to gain a better perspective on the principal points of interest in recent research trends. 
\vspace{-3pt}
\subsection{Cyber-physical System Security}\vspace{-3pt}
Several studies place particular focus on CPS security. To ensure data sources are authentic and reliable, in 2018, Fu et al. proposed using blockchain in CPS for an information security risk evaluation system in~\cite{fu2018cps}. Later, in 2020, Wang et al. analyzed the CPS data storage's security risks and proposed to utilize an improved blockchain mechanism for securing the data in~\cite{wang2020blockchain}. As the traditional Merkle hash tree fails to batch add/delete, they proposed to use the combination of accumulator and Merkle hash tree for non-membership proof. Rathore et al. proposed a secure deep learning (DL) method in~\cite{rathore2020blockchain} with blockchain for ensuring the cybersecurity of next-generation IoT CPS where decentralized, secure DL operations are performed at the edge nodes. This method contributed for big data analysis of contemporary CPS by deploying DL operation at edge layer and configuring distributed DL in a blockchain environment 
to ensure secure decentralization.
Lastly, Maloney et al. designed a security automation system in~\cite{maloney2020cyber} to deal with the operational security tasks and managing the security of CPS without repetitive duty through the integration of blockchain. The authors claim that this system, built on an Ethereum network, effectively increases the security of the CPS devices fleet and reduces complexity. 

\vspace{-3pt}
\subsection{Cyber-physical System Control}\vspace{-3pt}
Control is an essential factor in CPS, which can be tuned up through blockchain. Tan et al. proposed a blockchain-based access control scheme for Cyber-Physical-Social System (CPSS) in~\cite{tan2020blockchain}, where a node's account address in the blockchain is utilized as the identification number for accessing the CPSS big data. For redefining and storing the access control permission of CPSS big data, blockchain is utilized, which secures the processes of authorization, access control, authorization revocation, and audit. 
Garamvolgyi et al. \cite{Garamvolgyi:2018} focused on the control of CPS with the use of smart contracts. They proposed an approach in which smart contracts are produced from behavioral models, namely Unified Modeling Language (UML) statecharts, to coordinate the use of CPS elements. While the approach can be extended to other platforms, they presented a proof of concept using Ethereum smart contracts. Afanasev et al. \cite{Afanasev:2018} considered the advantages and disadvantages of blockchain and smart contract for control, workflow event logging, and monitoring in a Cyber-Physical Production System network. They proposed a blockchain-based architecture and provided relevant use cases. 

\begin{figure*}[t]
    \begin{center}
     \subfigure[]
        {
        \label{fig:pie}
            \includegraphics[width=0.44\textwidth, keepaspectratio=true]{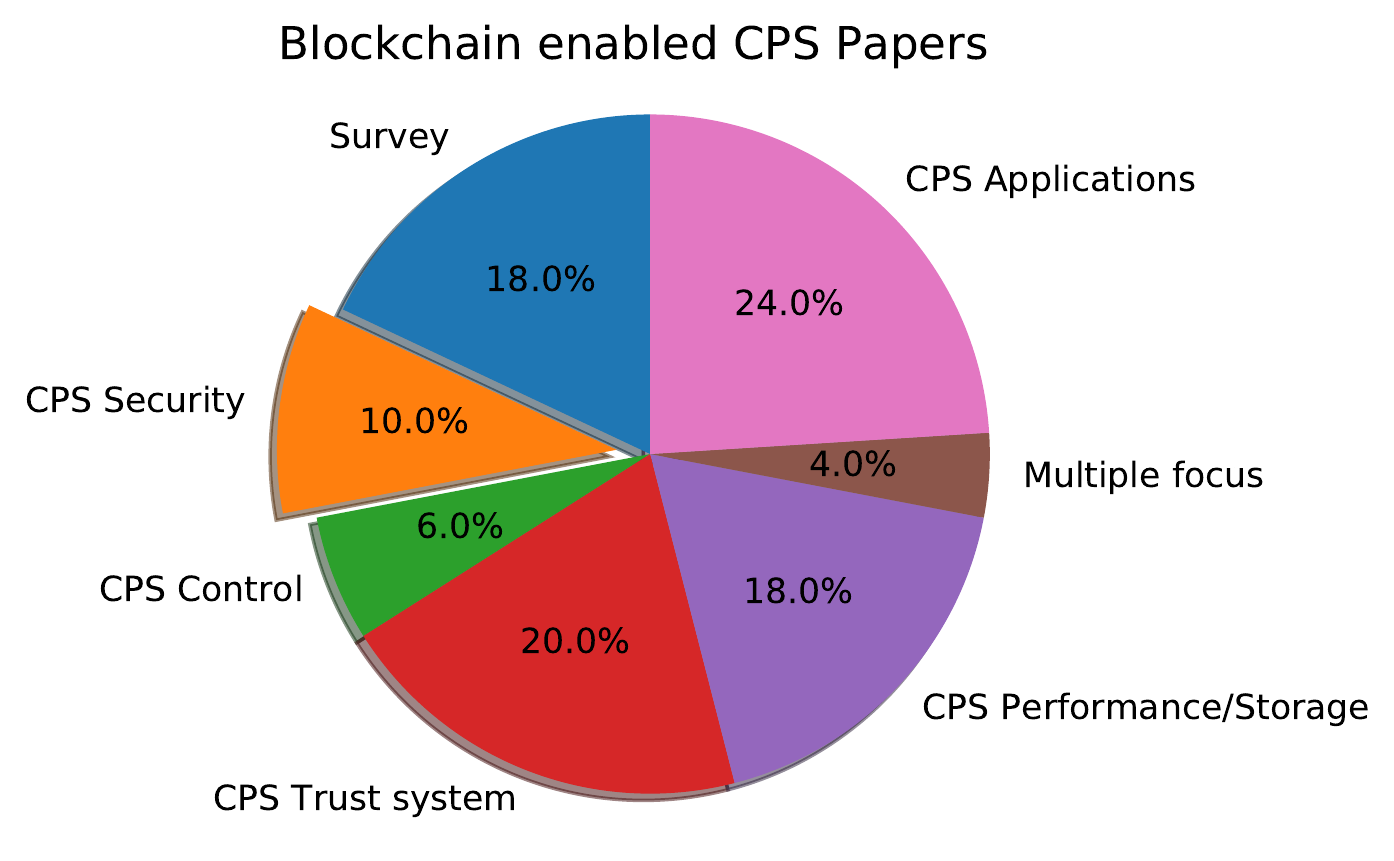}
        }     
    \subfigure[]
        {
        \label{fig:stacked}
            \includegraphics[width=0.40\textwidth, keepaspectratio=true]{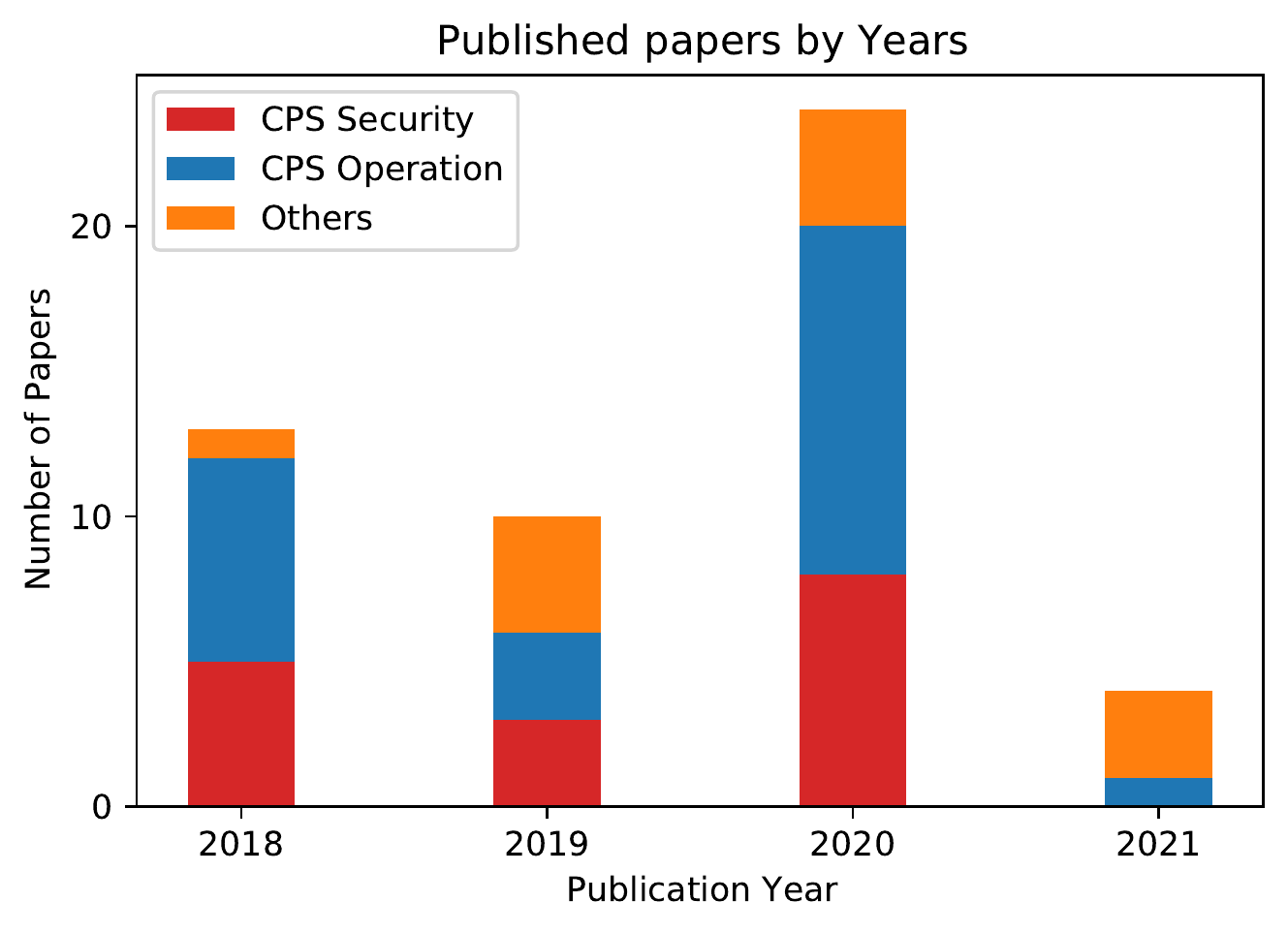}
        }     
      
    \end{center}
    \vspace{-15pt}
    \caption{Statistics of the papers published between 2018 to 2021, with (a) the pie chart according to the major concentration of the paper and (b) stacked bar chart of broad categorization of topics in each of the years.}
    \label{fig:stats}
    \vspace{-9pt}
\end{figure*}
\vspace{-4pt}
\subsection{Cyberphysical System Trust}\vspace{-3pt}
Given the importance of blockchain for establishing trust in CPS through decentralization and eliminating the middle man, it is no surprise that this research direction has been very active in the last few years. 

In 2018, Machado and Frohlich \cite{Machado:2018} presented a split blockchain-based architecture for increasing trust and decentralization for IoT data in CPS by using three levels to develop a chain of trust and using semi-trusted remote storage. 
Yang et al. \cite{Yang:2018} presented a method of decentralized private data acquisition blockchain using an on-demand data transmission routing algorithm and M/M/1/k queuing model to meet the trust and time consumption demands of CPPS. Afanasev et al. \cite{Afanasev2:2018} proposed the use of a blockchain network as a platform for a distributed decentralized network through the use of smart contracts for trustful communication between the nodes. 
While the authors identified several improvements that can be made, they presented the Ethereum blockchain as a positive alternative to current CPPS network alternatives. 
Also, Gries et al. \cite{Gries:2018} discussed the idea of using blockchain technology for scalable and decentralized trustful information flow tracking for CPS, using Information Flow Monitor (IFM) to visualize data without storing it. 

Later, in 2019, Kandah et al. \cite{Kandah:2019} presented a hardware-software co-design approach that includes RF-DNA fingerprinting for devices to have unique identities, behavioral trust management, a multi-layer decentralized database to manage trust information, and construction of a dynamic trust through RF-DNA fingerprinting,  and trust algorithms. 
Liu et al. \cite{Liu:2019} presented a blockchain-based technique to allow secure routing for Unmanned Aerial Systems (UAS) in mesh networks. The proposed strategy establishes trust through encryption and then uses blockchain to collect and redistribute routing information. One critical issue which their strategy addresses is that it enables source routing without revealing the mesh network's topology. In addition, LV et al. \cite{Lv:2019} proposed a blockchain-based publish/subscribe model for privacy in communication between sensing devices and interested nodes in the network. The proposed model sought to solve the trust problem, the issue of single-point failure, and uses public-key encryption with equality test. The authors noted that the use of ElGamal public-key cryptosystem with IND-CPA security ensures the confidentiality of the communications, while the use of the Ethereum ensures anonymity for publishers and subscribers. 

Then in 2020, Mohanta et al. proposed a signature storage solution for a diverse set of blockchain-based CPS applications for ensuring trust among the participating nodes in~\cite{mohanta2020trust}. The solution, built with Docker tools and Ethereum network, guarantees not only security properties but also reduces storage space and cost. 
Beckmann et al. proposed to use blockchain as the trust-enabling system component for  Cyber-Physical Trust Systems (CPTS), which is a CPS with IoT enriched with trust as a system component in~\cite{beckmann2020blockchain}. 
Milne et al. further elaborated the CPTS driven by blockchain in~\cite{milne2020cyber} by providing formal proofs of properties like integrity, identification, authentication, and non-repudiation using the Tamarin Prover tool.

\vspace{-2pt}
\subsection{Cyberphysical System Performance and/or Storage}\vspace{-3pt} 

Studies have also been carried out with a focus on solving the performance and storage issues caused by the exponentially growing number of devices and data for  CPS systems.

In 2018, Koumidis et al. \cite{Koumidis:2018} considered the integrity of CPS record logs for accountability and proposed a blockchain-based approach for computing block resource optimization in the PoW mechanism, including  computational cost.

In 2019, Li et al. \cite{Li:2019} proposed a blockchain dividing strategy using the community structure clustering method to decrease communication load, storage of dispensable data, and synchronization time. The proposed system also seeks to improve the concurrency of the system, as well as the efficiency of communication and data processing. 
Koumidis et al. \cite{Koumidis:2019} developed a blockchain technique for securing event logs in CPS, which bundles event data into blocks and delivers them to the system components that monitor and control the CPS in order to minimize the computational resources. 
Also, in \cite{Masood2:2019}, and \cite{Masood:2019}, Masood et al. presented a framework for a blockchain-based distributed management system for closed-loop CPS in order to address issues caused by computational constraints, centralized control, and network dependency. 

Then in 2020, Bouachir et al.~\cite{bouachir2020blockchain} presented an analysis of a fog-computing-based ecosystem integrated with blockchain for IIoT in order to manage and enhance the data storage, quality of service, and security requirements. 
In~\cite{fan2020blockchain}, for license-free spectrum resource management in Cyber-Physical-Social Systems (CPSSs), Fan et al. proposed a standard framework using blockchain and smart contracts that can be used for the edge computation of non-real-time data. For improving the overall transaction speed, they proposed a blockchain-KM protocol that effectively avoids losing typical attributes of a general blockchain. Also, Isaja et al. reported on FAR-EDGE experimentation of Smart Contracts and Blockchain, which proposed a reference architecture based on edge computing concentrating on efficient distributed computing power and network bandwidth usage, in~\cite{isaja2020blockchain}. 

Most recently, in 2021, Wang et al. \cite{Wang:2021} focused on the storage and computing challenges caused by IoT devices used with CPS cloud/edge computing. They proposed a Blockchain Software-Defined CPS (SD-CPS) framework that applies distributed resource management using cloud and edge computing to reduce system delay. 


\vspace{-3pt}
\subsection{Multiple Focus}\vspace{-3pt}

A few studies from 2021 emphasize a combination of objectives, including security, control, performance, data storage, and privacy. 
Neelam and Shinray \cite{Neelam:2021} presented an IoT-enabled CPS model using a fully programmable recursive internetworking architecture (RINA) with secure authentication using RINA password authentication for improving SDN and blockchain-enabled security. 
Rathore and Park \cite{Rathore:2021} addressed the challenges of centralized control, privacy, and security in deep learning (DL) for CPS. They proposed DeepBlockIoTNet, a DL approach for use in IoT CPS networks that uses blockchain for DL operations applied at the edge layer for decentralized and secure operations. 

\vspace{-3pt}
\subsection{Cyberphysical System Applications}\vspace{-3pt}

Finally, other studies focus on a variety of particular applications for blockchain in CPS, including shared manufacturing, smart grid, energy systems, intelligent robots, and Smart Controlled Business Environments (SCBE).  

In 2018, Zhao et al. \cite{Zhao:2018} addressed the issues of security and reliability of data distribution services. They proposed a secure pub-sub (SPS) architecture for blockchain-based fair payments with reputation, implemented with smart contracts and Ethereum network, which effectively eliminated the need for a reliable third party while maintaining confidentiality, the anonymity of the subscriber, and fairness. Wagner and McMillin \cite{Wagner:2018} considered security for VANETs, and presented a blockchain architecture with physically verified transactions, as well as a protocol for VANET security that does not require assistance from roadside units (RSUs). 
Teslya and Smirnov \cite{Teslya:2018} proposed a cyber-physical framework for the creation of intelligent robots that are considered separate entities, interacting with each other. This framework can also unite in a coalition to solve a common, complex problem with the help of blockchain technology with smart contracts. Lastly, Dong et al. \cite{Dong:2018} considered the opportunities and challenges presented by blockchain in uses for developments in energy systems and presented a prototype for future grids, which includes IoT, cloud, and blockchain.

\vspace{-1pt}
Then, in 2019, Patsonakis et al. \cite{Patsonakis:2019} also focused on energy systems, proposing a Demand Response (DR) energy system design that uses blockchains and smart contracts for decentralization to ensure security, privacy, reliability, audibility, and resistance to tampering. Gu et al. \cite{Gu:2019} presented a blockchain-based CPS security and safety protection framework for intelligent manufacturing CPS. They proposed that blockchain's distributed architecture can be used to optimize CPS layout and carry out data traceability while meeting the CPS safety requirements and even improving CPS safety through implementing the characteristics of data deposit and smart contract into CPS. Ahmadi-Assalemi et al. \cite{Ahmadi:2019} presented a framework using federated Blockchain (BC) model with a digital Chain-of-Custody (CoC) and a collaborative environment for the CPSs to serve as Digital Witnesses (DW) for investigations when an incident occurs. The framework facilitates object behavior tracking in Smart Controlled Business Environments (SCBE) and allows for proactive detection of insider threats. 

\vspace{-2pt}
In 2020, Kim et al. presented a comprehensive overview of the cyber-physical security vulnerabilities of the battery management system (BMS) from potential cyber-attacks in~\cite{kim2020overview}.In~\cite{vatankhah2020blockchain}, Barenji et al. addressed the security, scalability, and big-data problems for small and medium manufacturing enterprises (SMEs) by proposing a blockchain-based platform as a trustable network. This platform is built on a consortium blockchain which improves the consensus and communication protocols based on blockchain-enabled CPS. Yu et al. also addressed manufacturing and proposed a Blockchain-based Shared Manufacturing (BSM) framework for CPS based application support in~\cite{yu2020blockchain}, where the core operations are performed through a Resource Operation Blockchain (ROB), carrying out the basis of a consensus mechanism as well as a Smart Contract Network.
In~\cite{moore2020blockgrid}, Moore et al. presented the design and prototype of a blockchain implementation with CPS that consisted of a cluster of microcomputers forming a smart grid. These microcomputers, acting as nodes, are controlled by the smart contracts of a private blockchain.
Also, Shu et al. presented a two-layer model for  Medical CPSs (MCPS) in~\cite{shu2020efficient}, where medical records are stored off-blockchain and shared on-blockchain. They also proposed a certificate-less aggregate signature based on a multi-trapdoor hash function for MCPS. They claimed that because of avoiding exponential operations and bilinear maps, the proposed method is highly computationally efficient.  They further discussed the defense strategies leveraging blockchain technology in BMS, which can be used as the cybersecurity baseline reference. Also related to healthcare, in 2021, Rachakonda et al. \cite{Rachakonda:2021}  proposed Smart-Yoga Pillow, a Healthcare  CPS edge device that analyzes sleeping habits and physiological changes that occur during sleep, with a focus on the security of data transfer using RSA encryption, Ethereum blockchain, and access policy smart contracts.


\vspace{-4pt}
\section{Discussion}\vspace{-3pt}
\label{sec:discussion}

This section provides a statistical analysis of the papers published from the year 2018 to 2021 in blockchain-enabled CPS security and operation domains. Fig.~\ref{fig:stats} represents the statistical findings of the related publications, specifically Fig.~\ref{fig:pie} illustrates the pie chart of publications according to the major contribution of the paper, and Fig.~\ref{fig:stacked} represents the papers according to the publication year. 
From the pie chart, it is observed the largest group of the publications are application-oriented, which holds 24\% of the publications, followed closely by CPS Trust systems. This research illustrates the wide variety of blockchain applications for CPS, as well as the importance of blockchain in facilitating trust in CPS. The CPS security papers are holding close to one-tenth of the publications in these years. From the stacked bar chart, it is seen that while there was a continuation of CPS security concentrated papers till 2020, as the amount of CPS blockchain research grew significantly, the trend was shifting from the CPS security concentrated papers to the CPS operation focused papers. Also, there is a rising number of papers in the blockchain-enabled domains other than CPS security and operations. While the number of published papers is currently much lower in 2021 than in recent years, it is important to note that this survey only takes into account papers published through the beginning of June 2021. 

We also found some interesting trends from reviewing the literature. Among the research studies that propose blockchain-based models for CPS, 24 papers utilized smart contracts, which are a set of agreed-upon rules or terms that run on the blockchain to automate the execution of the terms without the need for a third party \cite{Masood2:2019}.  Of these 24 works, 16 used Ethereum as the open-source ledger platform for smart contracts. Also, 8 of the proposed models use edge networking, and 11 of the models leveraged encryption-based methods specifying the type. A detailed list of related reference papers is presented in Table.~\ref{tab:researchTrend}. Another insight is, for efficiency purposes, PoW like consensus mechanisms are too complex for CPS/IoT-based applications, leading to high delays and low throughput~\cite{braeken2020blockchain}. A topic worth considering in future studies on CPS-based usage of blockchain is the inclusion of greater mining incentives~\cite{braeken2020blockchain}. 

\begin{table}[hbt!]
\vspace{-5pt}
\caption{Research trend in Blockchain-enabled CPS}
\label{tab:researchTrend}
\begin{tabular}{|p{1.2cm}|p{1.2cm}|p{5.2cm}|}
\hline
\textbf{Aspect}                     & \textbf{Trend} & \textbf{Reference Paper}                                                                                                                                                                                                                                                                                                                                                                                                                                                                                                                                                                                                                                                                                                                                                                                                                                                                                                                                                                                                                       \\ \hline
\multirow{3}{*}{\textit{Technology}} & Smart Contract      & \begin{tabular}[c]{@{}l@{}}\cite{braeken2020blockchain}, \cite{wang2020blockchain}, \cite{maloney2020cyber}, \cite{Garamvolgyi:2018}, \cite{Afanasev:2018},  \cite{Machado:2018}, \cite{Afanasev2:2018}, \cite{Gries:2018}, \\ \cite{Liu:2019}, \cite{Lv:2019}, \cite{mohanta2020trust}, \cite{beckmann2020blockchain}, \cite{Masood2:2019}, \cite{Masood:2019}, \cite{fan2020blockchain}, \cite{isaja2020blockchain}, \\ \cite{Zhao:2018},  \cite{Teslya:2018}, \cite{Dong:2018}, \cite{Patsonakis:2019}, \cite{Ahmadi:2019}, \cite{yu2020blockchain}, \cite{moore2020blockgrid}, \cite{Rachakonda:2021}\end{tabular} \\ \cline{2-3} 
                                     & Encryption          & \begin{tabular}[c]{@{}l@{}}Asymmetric \cite{maloney2020cyber},  \cite{Yang:2018}, \cite{Liu:2019}, \cite{beckmann2020blockchain}, \cite{Rathore:2021}, \cite{Dong:2018}, \\ Public key \cite{Lv:2019}, \cite{Rathore:2021} \\ Advanced Encryption Standard (AES) \cite{Machado:2018}, \\ RSA Key \& Encryption/Decryption \cite{Wagner:2018}, \\ ElGamal \cite{Zhao:2018}, \\ Symmetric \cite{Liu:2019}\end{tabular}                                                                                                                                                                                                                                                                                                                                                                                         \\ \cline{2-3} 
                                     & Edge Net        & \cite{Li:2019}, \cite{Masood:2019}, \cite{fan2020blockchain}, \cite{isaja2020blockchain},  \cite{Wang:2021}, \cite{Rathore:2021}, \cite{vatankhah2020blockchain}, \cite{Rachakonda:2021}                                                                                                                                                                                                                                                                                                                                                                                                                                                                                                                                                                                                                                                                            \\ \hline
\multirow{3}{*}{\textit{Platform}}   & Ethereum            & \begin{tabular}[c]{@{}l@{}}\cite{maloney2020cyber}, \cite{Garamvolgyi:2018}, \cite{Afanasev:2018},  \cite{Machado:2018},  \cite{Afanasev2:2018}, \cite{Gries:2018}, \cite{Lv:2019}, \cite{mohanta2020trust},  \\ \cite{Masood:2019}, \cite{fan2020blockchain}, \cite{isaja2020blockchain}, \cite{Rathore:2021}, \cite{Zhao:2018}, \cite{yu2020blockchain}, \cite{moore2020blockgrid},  \cite{Rachakonda:2021}\end{tabular}                                                                                                                                                                                                                                                                                                \\ \cline{2-3} 
                                     & Bitcoin             & \cite{Zhao:2018}, \cite{Wagner:2018}                                                                                                                                                                                                                                                                                                                                                                                                                                                                                                                                                                                                                                                                                                                                                                                                                                                                                                                                                                         \\ \cline{2-3} 
                                     & EOS                 & \cite{tan2020blockchain}                                                                                                                                                                                                                                                                                                                                                                                                                                                                                                                                                                                                                                                                                                                                                                                                                                                                                                                                                                                                      \\ \hline
\end{tabular}
\end{table}


\vspace{-3pt}
\section{Conclusion}\vspace{-3pt}
\label{sec:conclusion}

The intrinsic combination of distributed data storage, consensus methods, and secure protocol implementations in blockchain efficiently solves diverse CPS performance and security issues. In this paper, we review current research on blockchain-enabled CPSs from both the security and operational viewpoints. In addition, we present some graphical representations of research works that summarize existing studies in an organized manner, which will aid future researchers in focusing on less explored areas.


\bibliographystyle{IEEEtran}

\bibliography{References}

\begin{thebibliography}{10}
\providecommand{\url}[1]{#1}
\csname url@samestyle\endcsname
\providecommand{\newblock}{\relax}
\providecommand{\bibinfo}[2]{#2}
\providecommand{\BIBentrySTDinterwordspacing}{\spaceskip=0pt\relax}
\providecommand{\BIBentryALTinterwordstretchfactor}{4}
\providecommand{\BIBentryALTinterwordspacing}{\spaceskip=\fontdimen2\font plus
\BIBentryALTinterwordstretchfactor\fontdimen3\font minus
  \fontdimen4\font\relax}
\providecommand{\BIBforeignlanguage}[2]{{%
\expandafter\ifx\csname l@#1\endcsname\relax
\typeout{** WARNING: IEEEtran.bst: No hyphenation pattern has been}%
\typeout{** loaded for the language `#1'. Using the pattern for}%
\typeout{** the default language instead.}%
\else
\language=\csname l@#1\endcsname
\fi
#2}}
\providecommand{\BIBdecl}{\relax}
\BIBdecl

\bibitem{Padilla:2021}
{R. Padilla, J. Sergent, J. Loehrke and G. Petras,}, ``Colonial pipeline
  reopens pipeline amid surge in gas shortages, higher gas prices and panic
  buying,''
  https://www.usatoday.com/in-depth/graphics/2021/05/12/colonial-pipeline-gas-shortage-prices-explained/5053043001/,
  [Online; accessed 14-Jun-2021].

\bibitem{Cisco:2020}
\BIBentryALTinterwordspacing
Cisco, ``Cisco annual internet report (2018–2023) white paper,'' 2020.
  [Online]. Available:
  \url{https://www.cisco.com/c/en/us/solutions/collateral/executive-perspectives/annual-internet-report/white-paper-c11-741490.pdf}
\BIBentrySTDinterwordspacing

\bibitem{casino2019systematic}
F.~Casino, T.~K. Dasaklis, and C.~Patsakis, ``A systematic literature review of
  blockchain-based applications: current status, classification and open
  issues,'' \emph{Telematics and informatics}, vol.~36, pp. 55--81, 2019.

\bibitem{Yaacoub:2020}
J.-P.~A. Yaacoub, O.~Salman, H.~N. Noura, N.~Kaaniche, A.~Chehab, and M.~Malli,
  ``Cyber-physical systems security: Limitations, issues and future trends,''
  \emph{Microprocessors and Microsystems}, 2020.

\bibitem{Shahid:2019}
A.~R. Shahid, N.~Pissinou, C.~Staier, and R.~Kwan, ``Sensor-chain: A
  lightweight scalable blockchain framework for internet of things,'' in
  \emph{2019 iThings and IEEE GreenCom-CPSCom-SmartData}, 2019.

\bibitem{Zhao:2021}
W.~Zhao, C.~Jiang, H.~Gao, S.~Yang, and X.~Luo, ``Blockchain-enabled
  cyber–physical systems: A review,'' \emph{IEEE IoT Journal}, 2021.

\bibitem{taylor2020systematic}
P.~J. Taylor, T.~Dargahi, A.~Dehghantanha, R.~M. Parizi, and K.-K.~R. Choo, ``A
  systematic literature review of blockchain cyber security,'' \emph{Digital
  Communications and Networks}, vol.~6, no.~2, pp. 147--156, 2020.

\bibitem{gupta2020smart}
R.~Gupta, S.~Tanwar, F.~Al-Turjman, P.~Italiya, A.~Nauman, and S.~W. Kim,
  ``Smart contract privacy protection using ai in cyber-physical systems:
  Tools, techniques and challenges,'' \emph{IEEE Access}, vol.~8, 2020.

\bibitem{Keshk:2021}
M.~Keshk, B.~Turnbull, E.~Sitnikova, D.~Vatsalan, and N.~Moustafa,
  ``Privacy-preserving schemes for safeguarding heterogeneous data sources in
  cyber-physical systems,'' \emph{IEEE Access}, vol.~9, 2021.

\bibitem{kanhere2020keynote}
S.~Kanhere, ``Keynote speech: Blockchain for cyber physical systems,'' in
  \emph{IEEE 2nd Internation Conference on BCCA}, 2020, pp. 1--1.

\bibitem{braeken2020blockchain}
A.~Braeken, M.~Liyanage, S.~S. Kanhere, and S.~Dixit, ``Blockchain and
  cyberphysical systems,'' \emph{IEEE Annals of the History of Computing},
  vol.~53, no.~09, pp. 31--35, 2020.

\bibitem{bodkhe2020survey}
U.~Bodkhe, D.~Mehta, S.~Tanwar, P.~Bhattacharya, P.~K. Singh, and W.-C. Hong,
  ``A survey on decentralized consensus mechanisms for cyber physical
  systems,'' \emph{IEEE Access}, vol.~8, pp. 54\,371--54\,401, 2020.

\bibitem{rathore2020survey}
H.~Rathore, A.~Mohamed, and M.~Guizani, ``A survey of blockchain enabled
  cyber-physical systems,'' \emph{Sensors}, vol.~20, no.~1, p. 282, 2020.

\bibitem{dedeoglu2020journey}
V.~Dedeoglu, A.~Dorri, R.~Jurdak, R.~A. Michelin, R.~C. Lunardi, S.~S. Kanhere,
  and A.~F. Zorzo, ``A journey in applying blockchain for cyberphysical
  systems,'' in \emph{IEEE COMSNETS}, 2020, pp. 383--390.

\bibitem{fu2018cps}
Y.~Fu, J.~Zhu, and S.~Gao, ``Cps information security risk evaluation based on
  blockchain and big data,'' \emph{Tehni{\v{c}}ki vjesnik}, vol.~25, 2018.

\bibitem{wang2020blockchain}
J.~Wang, W.~Chen, Y.~Ren, O.~Alfarraj, and L.~Wang, ``Blockchain based data
  storage mechanism in cyber physical system,'' \emph{Journal of Internet
  Technology}, vol.~21, no.~6, pp. 1681--1689, 2020.

\bibitem{rathore2020blockchain}
S.~Rathore and J.~H. Park, ``A blockchain-based deep learning approach for
  cyber security in next generation industrial cyber-physical systems,''
  \emph{IEEE Transactions on Industrial Informatics}, 2020.

\bibitem{maloney2020cyber}
M.~Maloney, G.~Falco, and M.~Siegel, ``Cyber-physical system security
  automation through blockchain remediation and execution (sabre).''

\bibitem{tan2020blockchain}
L.~Tan, N.~Shi, C.~Yang, and K.~Yu, ``A blockchain-based access control
  framework for cyber-physical-social system big data,'' \emph{IEEE Access},
  vol.~8, pp. 77\,215--77\,226, 2020.

\bibitem{Garamvolgyi:2018}
P.~Garamvölgyi, I.~Kocsis, B.~Gehl, and A.~Klenik, ``Towards model-driven
  engineering of smart contracts for cyber-physical systems,'' in
  \emph{IEEE/IFIP DSN-Workshop}, 2018, pp. 134--139.

\bibitem{Afanasev:2018}
M.~Y. Afanasev, Y.~V. Fedosov, A.~A. Krylova, and S.~A. Shorokhov, ``An
  application of blockchain and smart contracts for machine-to-machine
  communications in cyber-physical production systems,'' in \emph{2018 IEEE
  Industrial Cyber-Physical Systems (ICPS)}, 2018, pp. 13--19.

\bibitem{Machado:2018}
C.~Machado and A.~A. M.~Fröhlich, ``Iot data integrity verification for
  cyber-physical systems using blockchain,'' in \emph{IEEE ISORC}, 2018.

\bibitem{Yang:2018}
T.~Yang, F.~Zhai, J.~Liu, M.~Wang, and H.~Pen, ``Self-organized cyber physical
  power system blockchain architecture and protocol,'' \emph{International
  Journal of Distributed Sensor Networks}, vol.~14, 2018.

\bibitem{Afanasev2:2018}
M.~Y. Afanasev, A.~A. Krylova, S.~A. Shorokhov, Y.~V. Fedosov, and A.~S.
  Sidorenko, ``A design of cyber-physical production system prototype based on
  an ethereum private network,'' in \emph{2018 22nd Conference of Open
  Innovations Association (FRUCT)}, 2018, pp. 3--11.

\bibitem{Gries:2018}
S.~Gries, O.~Meyer, F.~Wessling, M.~Hesenius, and V.~Gruhn, ``Using blockchain
  technology to ensure trustful information flow monitoring in cps,'' in
  \emph{2018 IEEE International Conference on Software Architecture Companion
  (ICSA-C)}, 2018, pp. 35--38.

\bibitem{Kandah:2019}
F.~Kandah, J.~Cancelleri, D.~Reising, A.~Altarawneh, and A.~Skjellum, ``A
  hardware-software codesign approach to identity, trust, and resilience for
  iot/cps at scale,'' in \emph{2019 International Conference on Internet of
  Things (iThings) and IEEE Green Computing and Communications (GreenCom) and
  IEEE Cyber, Physical and Social Computing (CPSCom) and IEEE Smart Data
  (SmartData)}, 2019, pp. 1125--1134.

\bibitem{Liu:2019}
Y.~Liu, J.~Wang, H.~Song, J.~Li, and J.~Yuan, ``Blockchain-based secure routing
  strategy for airborne mesh networks,'' in \emph{2019 IEEE International
  Conference on Industrial Internet (ICII)}, 2019.

\bibitem{Lv:2019}
P.~Lv, L.~Wang, H.~Zhu, W.~Deng, and L.~Gu, ``An iot-oriented
  privacy-preserving publish/subscribe model over blockchains,'' \emph{IEEE
  Access}, vol.~7, pp. 41\,309--41\,314, 2019.

\bibitem{mohanta2020trust}
B.~K. Mohanta, U.~Satapathy, M.~R. Dey, S.~S. Panda, and D.~Jena, ``Trust
  management in cyber physical system using blockchain,'' in \emph{2020 11th
  International Conference on Computing, Communication and Networking
  Technologies (ICCCNT)}.\hskip 1em plus 0.5em minus 0.4em\relax IEEE, 2020,
  pp. 1--5.

\bibitem{beckmann2020blockchain}
A.~Beckmann, A.~Milne, J.-J. Razafindrakoto, P.~Kumar, M.~Breach, and
  N.~Preining, ``Blockchain-based cyber physical trust systems,'' \emph{IoT
  Security: Advances in Authentication}, pp. 265--277, 2020.

\bibitem{milne2020cyber}
A.~J. Milne, A.~Beckmann, and P.~Kumar, ``Cyber-physical trust systems driven
  by blockchain,'' \emph{IEEE Access}, vol.~8, pp. 66\,423--66\,437, 2020.

\bibitem{Koumidis:2018}
K.~Koumidis, P.~Kolios, and C.~Panayiotou, ``Optimizing blockchain for data
  integrity in cyber physical systems,'' 08 2018, pp. 73--80.

\bibitem{Li:2019}
S.~Li, H.~Xiao, H.~Wang, T.~Wang, J.~Qiao, and S.~Liu, ``Blockchain dividing
  based on node community clustering in intelligent manufacturing cps,'' in
  \emph{2019 IEEE International Conference on Blockchain (Blockchain)}, 2019,
  pp. 124--131.

\bibitem{Koumidis:2019}
K.~Koumidis, P.~Kolios, G.~Ellinas, and C.~G. Panayiotou, ``Secure event
  logging using a blockchain of heterogeneous computing resources,'' in
  \emph{2019 IEEE Global Communications Conference (GLOBECOM)}, 2019.

\bibitem{Masood2:2019}
A.~B. Masood, M.~Lestas, H.~K. Qureshi, N.~Christofides, N.~Ashraf, and
  F.~Mehmood, ``Closing the loop in cyber-physical systems using blockchain:
  Microgrid frequency control example,'' in \emph{2019 2nd IEEE Middle East and
  North Africa COMMunications Conference (MENACOMM)}, 2019, pp. 1--6.

\bibitem{Masood:2019}
A.~Bin~Masood, H.~K. Qureshi, S.~M. Danish, and M.~Lestas, ``Realizing an
  implementation platform for closed loop cyber-physical systems using
  blockchain,'' in \emph{IEEE VTC2019-Spring}, 2019.

\bibitem{bouachir2020blockchain}
O.~Bouachir, M.~Aloqaily, L.~Tseng, and A.~Boukerche, ``Blockchain and fog
  computing for cyberphysical systems: The case of smart industry,''
  \emph{Computer}, vol.~53, no.~9, pp. 36--45, 2020.

\bibitem{fan2020blockchain}
X.~Fan and Y.~Huo, ``Blockchain based dynamic spectrum access of non-real-time
  data in cyber-physical-social systems,'' \emph{IEEE Access}, 2020.

\bibitem{isaja2020blockchain}
M.~Isaja and A.~Cal, ``Blockchain as a key enabling technology for
  decentralized cyber-physical production systems,'' 2020.

\bibitem{Wang:2021}
D.~Wang, N.~Zhao, B.~Song, P.~Lin, and F.~R. Yu, ``Resource management for
  secure computation offloading in softwarized cyber–physical systems,''
  \emph{IEEE Internet of Things Journal}, 2021.

\bibitem{Neelam:2021}
B.~S. Neelam and B.~A. Shimray, ``Applicability of rina in iot communication
  for acceptable latency and resiliency against device authentication
  attacks,'' in \emph{2021 6th International Conference for Convergence in
  Technology (I2CT)}, 2021, pp. 1--7.

\bibitem{Rathore:2021}
S.~Rathore and J.~H. Park, ``A blockchain-based deep learning approach for
  cyber security in next generation industrial cyber-physical systems,''
  \emph{IEEE Transactions on Industrial Informatics}, vol.~17, 2021.

\bibitem{Zhao:2018}
Y.~Zhao, Y.~Li, Q.~Mu, B.~Yang, and Y.~Yu, ``Secure pub-sub: Blockchain-based
  fair payment with reputation for reliable cyber physical systems,''
  \emph{IEEE Access}, vol.~6, pp. 12\,295--12\,303, 2018.

\bibitem{Wagner:2018}
M.~Wagner and B.~McMillin, ``Cyber-physical transactions: A method for securing
  vanets with blockchains,'' in \emph{2018 IEEE 23rd Pacific Rim International
  Symposium on Dependable Computing (PRDC)}, 2018.

\bibitem{Teslya:2018}
N.~Teslya and A.~Smirnov, ``Blockchain-based framework for ontology-oriented
  robots’ coalition formation in cyberphysical systems,'' \emph{MATEC Web of
  Conferences}, vol. 161, p. 03018, 01 2018.

\bibitem{Dong:2018}
Z.~Dong, F.~Luo, and G.~Liang, ``Blockchain: a secure, decentralized, trusted
  cyber infrastructure solution for future energy systems,'' \emph{Journal of
  Modern Power Systems and Clean Energy}, vol.~6, 2018.

\bibitem{Patsonakis:2019}
C.~Patsonakis, S.~Terzi, I.~Moschos, D.~Ioannidis, K.~Votis, and D.~Tzovaras,
  ``Permissioned blockchains and virtual nodes for reinforcing trust between
  aggregators and prosumers in energy demand response scenarios,'' in
  \emph{IEEE EEEIC/ICPS Europe}, 2019.

\bibitem{Gu:2019}
A.~Gu, Z.~Yin, C.~Fan, and F.~Xu, ``Safety framework based on blockchain for
  intelligent manufacturing cyber physical system,'' in \emph{1st International
  Conference on Industrial Artificial Intelligence (IAI)}, 2019.

\bibitem{Ahmadi:2019}
G.~Ahmadi-Assalemi, H.~M. al~Khateeb, G.~Epiphaniou, J.~Cosson, H.~Jahankhani,
  and P.~Pillai, ``Federated blockchain-based tracking and liability
  attribution framework for employees and cyber-physical objects in a smart
  workplace,'' in \emph{2019 IEEE 12th International Conference on Global
  Security, Safety and Sustainability (ICGS3)}, 2019, pp. 1--9.

\bibitem{kim2020overview}
T.~Kim, J.~Ochoa, T.~Faika, A.~Mantooth, J.~Di, Q.~Li, and Y.~Lee, ``An
  overview of cyber-physical security of battery management systems and
  adoption of blockchain technology,'' \emph{IEEE Journal of Emerging and
  Selected Topics in Power Electronics}, 2020.

\bibitem{vatankhah2020blockchain}
A.~Vatankhah~Barenji, Z.~Li, W.~M. Wang, G.~Q. Huang, and D.~A. Guerra-Zubiaga,
  ``Blockchain-based ubiquitous manufacturing: A secure and reliable
  cyber-physical system,'' \emph{International Journal of Production Research},
  vol.~58, no.~7, pp. 2200--2221, 2020.

\bibitem{yu2020blockchain}
C.~Yu, X.~Jiang, S.~Yu, and C.~Yang, ``Blockchain-based shared manufacturing in
  support of cyber physical systems: concept, framework, and operation,''
  \emph{Robotics and Computer-Integrated Manufacturing}, 2020.

\bibitem{moore2020blockgrid}
G.~M. Moore, ``Blockgrid: A blockchain-mediated cyber-physical instructional
  platform,'' Naval Postgraduate School, Tech. Rep., 2020.

\bibitem{shu2020efficient}
H.~Shu, P.~Qi, Y.~Huang, F.~Chen, D.~Xie, and L.~Sun, ``An efficient
  certificateless aggregate signature scheme for blockchain-based medical cyber
  physical systems,'' \emph{Sensors}, vol.~20, no.~5, p. 1521, 2020.

\bibitem{Rachakonda:2021}
L.~Rachakonda, A.~K. Bapatla, S.~P. Mohanty, and E.~Kougianos, ``Sayopillow:
  Blockchain-integrated privacy-assured iomt framework for stress management
  considering sleeping habits,'' \emph{IEEE Transactions on Consumer
  Electronics}, vol.~67, no.~1, pp. 20--29, 2021.

\end{thebibliography}


\end{document}